\documentclass{article}
\usepackage{latexsym}
\usepackage{graphicx}
\begin{document}
\begin{center}
\textbf{\LARGE Classical and Relativistic Derivation of the \vspace{.3cm} \newline  Sagnac Effect}
\end{center}
\begin{large}
\begin{center}
W. Engelhardt\footnote{Home address: Fasaneriestrasse 8, D-80636 M\"{u}nchen, Germany\par \hspace*{.15cm} Electronic address: wolfgangw.engelhardt@t-online.de}, retired from:
\end{center}

\begin{center}
Max-Planck-Institut f\"{u}r Plasmaphysik, D-85741 Garching, Germany
\end{center}

\vspace{.6cm}

\noindent \textbf{Abstract}

\noindent Both the classical and the relativistic composition law for velocities are applied to re-calculate the Sagnac Effect. The ensuing formulae for the fringe shift are found to differ already in first order of v/c. Whilst the classical formula is validated by interferometric measurements and verified by the GPS-system, this is not the case for the relativistic result.

\vspace{.6cm}

\vspace{.5cm}
\noindent \textbf{\large P.A.C.S.:}  07.60.Ly, 42.81.Pa, 04.20.-q, 03.30.+p

\vspace{1.0cm}

\noindent \textbf{1 Introduction}

\noindent In 1925 Michelson and Gale built a huge earth-fixed Sagnac Interferometer in Illinois [1] demonstrating that the light velocity is anisotropic on the rotating earth. For Sagnac this result did not come as a surprise having explained the underlying effect on the basis of the ether theory in 1913 [2]. The Special Relativity Theory (SRT), however, had predicted on the basis of the Lorentz Transformation (LT) that the velocity of light is isotropic in all inertial systems [3]. Of course, the earth is not an inertial system, but due to the earth's rotation the motion of Michelson's device was only about 2 mm during the passage of a light beam round the interferometer. The deviation of the mirrors from a straight orbit was less than one hundredth of an atomic diameter so that non-inertial effects should be entirely negligible. Nevertheless, the linear term x v/c$^{2}$ in the LT -- which is responsible for c = const -- was simply not there. The experimental result was consistent with the Galilei Transformation (GT), i.e. light seemed to travel at the velocity c - v from west to east and c + v on the return path from east to west, where v is the local rotational velocity of the earth's surface. 

Strictly speaking, the rotating earth must be described by the General Relativity Theory, but it is hard to see how a very slow rotation should invalidate the LT and re-establish the GT as found by Michelson and Gale. Indeed, Post admits in his great review article [4]: ``The search for a physically meaningful transformation for rotation is not aided in any way whatever by the principle of general space-time covariance, nor is it true that the space-time theory of gravitation plays any role in establishing physically correct transformations.'' Consequently, he sticks to the classical formula valid in the ether theory making only a slight modification which is too small to be verified by experiment.

In this paper we re-derive the Sagnac Effect on the basis both of the classical and of the relativistic composition law of velocities that can also be formulated for a curved light path according to [5]. Calculating the ensuing travel times of light round the interferometer we find that the LT -- due to its linear term x v/c$^{2}$ -- does not predict any Sagnac Effect, but results in c = const also in a rotating system as it does in an inertial system. This explains then why Ashby [6], e.g., uses the Newtonian or Galilean time transformation t' = t rather than t' = $\gamma $ (t - x v/c$^{2})$ when he calculates the Sagnac Effect in the GPS-System. This was also observed by Carroll Alley in a comment at the end of an engineering presentation on \textit{GPS and Relativity} [7]. In Sect. 2 we derive the classical Sagnac Effect for a circular light beam, and in Sect. 3 we do the same making use of the LT. Sect. 4 discusses the implications of our results.

\vspace{.6cm}
\noindent\textbf{2 Sagnac Effect explained in the framework of the ether theory}

\noindent The classical transformation law between inertial systems is the GT which reads:
\begin{equation}
\label{eq1}
\vec {x}\,'=\vec {x}-\vec {v}\,t\,,\;t\,'=t
\end{equation}
Although it is formulated only for a linear constant velocity between the systems, it can also be adapted for a constant rotation by deriving a composition law for circular velocities. Considering infinitesimal line elements on a circular orbit one can write:
\begin{equation}
\label{eq2}
ds\,'=ds-v_0 \,dt\;,\quad dt\,'=dt
\end{equation}
where $v_0 =\Omega \,R$ is now the rotational velocity with angular frequency $\Omega $ on the circular orbit $s$ at radius $R$. Dividing by the time increment on both sides of equation (\ref{eq2}) one obtains with ${ds} \mathord{\left/ {\vphantom {{ds} {dt=v_\varphi }}} \right. \kern-\nulldelimiterspace} {dt=v_\varphi }$ the composition law for constant circular velocities:
\begin{equation}
\label{eq3}
v_\varphi '=v_\varphi -v_0 
\end{equation}

In fact, facilitating the analysis most treatments of the Sagnac Effect consider a circular interferometer where the light beam is guided on a circle either by a large amount of tangential mirrors or by optical fibres [4], [5]. The travel time of light to complete the orbit $L=2\pi \,R$ around the circumference of the interferometer is then
\begin{equation}
\label{eq4}
t^\pm =\frac{L}{v_\varphi \pm v_0 }=\frac{L}{c\pm v_0 }
\end{equation}
depending on whether the light propagates parallel or anti-parallel to the rotational velocity. Two coherent light beams starting at the beam splitter in opposite directions return at different times leading to a fringe shift due to the time difference as calculated with (\ref{eq4}):
\begin{equation}
\label{eq5}
\Delta t=\frac{L}{c-v_0 }-\frac{L}{c+v_0 }=\frac{2\,L\,v_0 }{c^2-v_0^2 }=\frac{2\,L\,v_0 }{c^2}\frac{1}{1-{v_0^2 } \mathord{\left/ {\vphantom {{v_0^2 } {c^2}}} \right. \kern-\nulldelimiterspace} {c^2}}
\end{equation}
The linear dependence on the rotational velocity has been confirmed by many experiments, whereas the quadratic term is too small to be measured in practice. In most textbooks result (\ref{eq5}) is expressed by the fringe shift in units of the wavelength $\lambda _0 $
\begin{equation}
\label{eq6}
\Delta Z=\frac{4\,\vec {A}\cdot \,\vec {\Omega }}{c\,\lambda _0 }
\end{equation}
where, in general, the scalar product of the oriented area $\vec {A}$ enclosed by the light path with the vector angular velocity $\vec {\Omega }$ enters. One can show that the fringe shift is independent of the shape of $\vec {A}$ and of the position of the rotational axis, but depends on the cosine of the angle between $\vec {A}\,\,\mbox{and}\,\,\vec {\Omega }$.

Whereas Sagnac could rotate his device in the laboratory at variable speed in order to verify formula (\ref{eq6}), Michelson and Gale had to vary the area enclosed by the light path in order to prove the validity of (\ref{eq6}). This experiment was important as it demonstrated that the ether is not co-rotating with the earth like air. Apparently, an ``ether wind'' must exist due to the rotation of the earth which leads to the observed time difference (\ref{eq5}). Any translational motion, which surely exists, cannot be detected by the circular interferometer, as the influence of a constant translational velocity cancels on the roundtrip of the light. In this respect the Sagnac Interferometer is as insensitive as the Michelson-Morley Interferometer [8] where the enclosed area $A$ vanishes.

Although the experimental results obtained by Sagnac Interferometers are in full agreement with the ether theory, as shown above, they are generally not considered as disproving the SRT which postulates c = const and is thus at variance with the ether composition law (\ref{eq3}). In the next Section we will show why it is commonly held that there is no discrepancy between the classical composition law (\ref{eq3}) and the relativistic philosophy concerning the behaviour of light. The reason is a defective application of SRT.

\vspace{.6cm}
\noindent \textbf{3 Sagnac Effect interpreted in the framework of the Special Relativity \newline Theory}

\noindent Most textbooks assure the reader that the Sagnac Effect is a ``relativistic'' effect which incidentally may also be derived by the ether theory leading practically to the same result. Looking closer one finds, however, that a truly relativistic derivation is not really attempted. A good example is Post's procedure in Sect. III [4]. He also uses the circular geometry (see Fig. 8) which our analysis in the previous Section was referring to. For the stationary observer he comes up with the time difference (7 P) which is identical with our formula (\ref{eq5}) derived from the classical composition law for velocities. This must also hold for the co-moving observer, if Ashby's [6] Galilean time transformation t' = t into the rotating system is valid. 

Post insists, however, that the time increment between the rotating and the stationary system must transform according to his equation (11 P):
\[
dt=\gamma \,dt\,'\,,\quad \gamma =1 \mathord{\left/ {\vphantom {1 {\sqrt {1-{v^2} \mathord{\left/ {\vphantom {{v^2} {c^2}}} \right. \kern-\nulldelimiterspace} {c^2}} }}} \right. \kern-\nulldelimiterspace} {\sqrt {1-{v^2} \mathord{\left/ {\vphantom {{v^2} {c^2}}} \right. \kern-\nulldelimiterspace} {c^2}} }
\]
which can also be found in Malykin's paper [5]. Consequently, these authors claim that the correct relativistic time difference as measured on the rotating beam splitter must be Post's equation (23 P) or Malykin's (5 M) which is our classical formula (\ref{eq5}) divided by the $\gamma $-factor. One notices that the LT was only applied in a mutilated form  when (23 P) was derived, since (11 P) is at variance with the complete LT formula. A different conclusion is reached by employing the correct expression, i.e. by applying the truly relativistic composition law for velocities that cannot be obtained from (11 P). 

For the case where the light beam propagates parallel to the $x$-axis the LT reads:
\begin{equation}
\label{eq7}
x'=\gamma \left( {x-\,v\,t} \right)\,,\quad t'=\gamma \left( {t-\,{x\,v} \mathord{\left/ {\vphantom {{x\,v} {c^2}}} \right. \kern-\nulldelimiterspace} {c^2}} \right)
\end{equation}
In order to adapt these formulae to a curved light path, we consider infinitesimal time increments in analogy to equation (\ref{eq2}) assuming a constant rotational velocity $v_0 $:
\begin{equation}
\label{eq8}
dt'=\gamma \left( {dt-\,{v_0 \,ds} \mathord{\left/ {\vphantom {{v_0 \,ds} {c^2}}} \right. \kern-\nulldelimiterspace} {c^2}} \right)
\end{equation}
where $s$ denotes the curved light path in the rotating interferometer. This formula can also be found in Post's article (29 P) where he quotes papers by Langevin and Trocheris. Of course, (\ref{eq8}) is only justified as long as the light beam propagates parallel to the rotational velocity, but this is ensured by the use of an optical fibre, for example. The spatial increment
\begin{equation}
\label{eq9}
ds'=\gamma \left( {ds-\,v_0 \,dt} \right)
\end{equation}
is not given by Post, nor does he calculate the composition rule resulting from division of (\ref{eq9}) by (\ref{eq8}):
\begin{equation}
\label{eq10}
v_\varphi '=\frac{v_\varphi -v_0 }{1-{v_\varphi \,v_0 } \mathord{\left/ {\vphantom {{v_\varphi \,v_0 } {c^2}}} \right. \kern-\nulldelimiterspace} {c^2}}
\end{equation}
Although not explicitly derived, this formula can be found in Malykin's paper [5] as the second equation (2 M). He does, however, not draw the obvious conclusion: If the phase velocity of the light in the laboratory is 
\begin{equation}
\label{eq11}
v_\varphi =c
\end{equation}
it follows from (\ref{eq10}) that the phase velocity in the rotating system is also
\begin{equation}
\label{eq12}
v_\varphi '=c
\end{equation}
This result does not really come as a surprise, since it only reflects Voigt's postulate $c=\mbox{const}$ from which the LT (\ref{eq7}) was derived [3]. If this transformation holds between inertial systems, it must also hold between an inertial system and a system rotating with constant velocity as just demonstrated.

The obvious consequence of Malykin's formula (2 M), or (\ref{eq10}) above is that coherent beams leaving the beam splitter at the same time in opposite directions will return at the same time as they both travel at the same speed $c$ according to (\ref{eq12}). The relativistically correct result is, therefore, not Post's formula (7 P), or Malykin´s formula (5 M), but simply
\begin{equation}
\label{eq13}
\Delta t'=t'{\kern 1pt}^+-t'{\kern 1pt}^-=L \mathord{\left/ {\vphantom {L c}} \right. \kern-\nulldelimiterspace} c-L \mathord{\left/ {\vphantom {L c}} \right. \kern-\nulldelimiterspace} c=0
\end{equation}
In other words, the SRT correctly applied to a rotating Sagnac Interferometer does not predict the Sagnac Effect. 

\vspace{.6cm}
\noindent \textbf{4 Implications of the demonstrated results}

\noindent The Michelson-Gale experiment of 1925 had already shown that Voigt's postulate c = const is untenable. When the GPS-system delivered more and more precise measurements based on the propagation of microwaves, it became clear that the Sagnac Effect had to be taken into account. This implied the use of the classical formulae derived in Sect. 2 on the basis of the ether theory. Ashby drew the correct conclusion and re-introduced the Galilean time transformation t' = t between the inertial ECI and the rotating ECEF system [6]. From Alley's remark at the end of [7] we conclude that the GPS-software -- developed by engineers for providing precise position measurements -- ignores the Lorentz time transformation, following Ashby's line instead. There are, of course, other indications voiced by Hatch [9] which suggest that the SRT should not be applied to GPS-measurements.

In the light of the analysis in Sect. 3 and the irrefutable experimental results it is obvious that Voigt's postulate of 1887 [10] does not reflect an observed physical law. It was always clear that it was invalid in the case of acoustics, and we have no reason to adopt it in optics either. As the SRT is a consequence of this postulate, its basic tenet is at stake.

\vspace{.6cm}

\end{large}

\vspace{.6cm}
\begin{thebibliography}{99}
\bibitem{Michelson}{Albert Abraham Michelson, Henry G. Gale: \textit{The Effect of the Earth's Rotation on the Velocity of Light II}. In: \textit{The Astrophysical Journal} \textbf{61} (1925) 140-145.}
\bibitem {Sagnac}{Georges Sagnac: \textit{Sur la preuve de la r\'{e}alit\'{e} de l'\'{e}ther lumineux par l'exp\'{e}rience de l'interf\'{e}rographe tournant}. In: \textit{Comptes Rendus} \textbf{157} (1913)~1410-1413.}
\bibitem{Engelhardt1}{ Wolfgang Engelhardt, \textit{On the Origin of the Lorentz Transformation}. In: \underline {http://arxiv.org/abs/1303.5309}. }
\bibitem{Post}{Evert Jan Post, \textit{Sagnac Effect. }In: \textit{Review of Modern Physics }\textbf{39} (1967) 475-493.} (\underline {http://www.orgonelab.org/EtherDrift/Post1967.pdf})
\bibitem{Malykin}G. B. Malykin, \textit{The Sagnac effect: correct and incorrect explanations}. In: \textit{Physics-Uspekhi} \textbf{43} (2000) 1229-1252.\\
See also equation (1.1) in:\\
{Annemarie Schied: \textit{Helium-Neon-Ringlaser-Gyroskop}. Staatsexamensarbeit, University of Ulm (2006) \newline \underline {http://www.uni-ulm.de/fileadmin/website{\_}uni{\_}ulm/nawi.inst.220/publikationen/}\newline \underline {LehramtAnnemarieSchied2006.pdf.}}
\bibitem{Ashby}{Neil Ashby: \textit{Relativity in the Global Positioning System}. In: \textit{Living Rev.Relativity} \textbf{6} (2003) 1. Sect. 2, eq. (\ref{eq3}) (\underline {http://www.livingreviews.org/lrr-2003-1}).}
\bibitem{Fliegel}{Henry F. Fliegel, Raymond S. DiEsposti: \textit{GPS and Relativity: An Engineering Overview. }In: \textit{28th Annual Precise Time and Time Interval (PTTI) Applications and Planning Meeting,} VA, 3-5 Dec 1996, pp. 189-199 (\underline {http://tycho.usno.navy.mil/ptti/1996/Vol{\%}2028{\_}16.pdf}).}
\bibitem{Engelhardt2}{Wolfgang Engelhardt, \textit{Phase and frequency shift in a Michelson Interferometer. }Submitted for publication in: \textit{Physics Essays} (2013).}
\bibitem{Hatch}{Ronald R. Hatch: \textit{Clock Behavior and the Search for an Underlying Mechanism for Relativistic Phenomena.} In: \textit{Proceedings of the 58th Annual Meeting of The Institute of Navigation and CIGTF 21st Guidance Test Symposium}, Albuquerque, NM, June 2002, pp. 70-81 
(\underline{https://www.ion.org/publications/abstract.cfm?articleID=937}).}
\bibitem{Voigt}{Woldemar Voigt: \textit{Ueber das Doppler'sche Princip. }In:\textit{ Nachrichten von der K\"{o}niglichen} \textit{Gesellschaft der Wissenschaften und der Georg--Augusts--Universit\"{a}t zu G\"{o}ttingen,} No. 2, 10. M\"{a}rz 1887.}
\end{thebibliography}
\end{document}